\documentclass[aps,prb,a4,twocolumn,superscriptaddress,showpacs]{revtex4}

\usepackage{times}
\usepackage{amssymb,amsmath}
\usepackage{graphicx}
\usepackage{natbib}
\usepackage{multirow}
\usepackage{color}
\usepackage[a4paper,left=2cm,top=3cm,right=1cm,bottom=2cm]{geometry}

\begin{document}

\title{Non-calorimetric determination of {\color{black}absorbed power} during magnetic nanoparticle based hyperthermia}

\author{I. Gresits}
\affiliation{Department of Non-Ionizing Radiation, National Public Health Institute, Budapest}

\author{Gy. Thur\'{o}czy}
\affiliation{Department of Non-Ionizing Radiation, National Public Health Institute, Budapest}

\author{O. S\'agi}
\affiliation{Department of Physics, Budapest University of Technology and Economics and MTA-BME Lend\"{u}let Spintronics Research Group (PROSPIN), Po. Box 91, H-1521 Budapest, Hungary}

\author{B. Gy\"ure-Garami}
\affiliation{Department of Physics, Budapest University of Technology and Economics and MTA-BME Lend\"{u}let Spintronics Research Group (PROSPIN), Po. Box 91, H-1521 Budapest, Hungary}

\author{B. G. M\'arkus}
\affiliation{Department of Physics, Budapest University of Technology and Economics and MTA-BME Lend\"{u}let Spintronics Research Group (PROSPIN), Po. Box 91, H-1521 Budapest, Hungary}

\author{F. Simon\email{f.simon@eik.bme.hu}}
\affiliation{Department of Physics, Budapest University of Technology and Economics and MTA-BME Lend\"{u}let Spintronics Research Group (PROSPIN), Po. Box 91, H-1521 Budapest, Hungary}

\begin{abstract}
Nanomagnetic hyperthermia (NMH) is intensively studied with the prospect of cancer therapy. A major challenge is to determine the dissipated power during \textit{in vivo} conditions and conventional methods are either invasive or inaccurate. We present a non-calorimetric method which yields the heat absorbed during hyperthermia: it is based on accurately measuring the quality factor change of a resonant radio frequency circuit which is employed for the irradiation. {\color{black}The approach provides the absorbed power in real-time, without the need to monitor the sample temperature as a function of time. As such, it is free from the problems caused by the non-adiabatic heating conditions of the usual calorimetry.} We validate the method by comparing the dissipated power with a conventional calorimetric measurement. We present the validation for two types of resonators with very different filling factors: a solenoid and a so-called birdcage coil. The latter is a volume coil, which is generally used in magnetic resonance imaging (MRI) under in vivo condition. The presented method therefore allows to effectively combine MRI and thermotherapy and is thus readily adaptable to existing imaging hardware.\end{abstract}

\maketitle

\section{Introduction}
Cancer is one of the major death causes worldwide, with several proven and being developed therapeutic methods \cite{BEIK2016205}. One promising cancer therapeutic method is hyperthermia. It involves raising the temperature of the local environment of a tumour, which results in an induced cell death\cite{{PankhurstReview2003},{PankhurstHyperthermia},{KUMAR2011789},{ANGELAKERIS20171642},{GIUSTINI2010},{KRISHNAN2010}} due to the large susceptibility of tumour cells to temperature as compared to healthy ones. In addition, local heating may enhance the efficiency of ionizing radiotherapy i.e. combining thermotherapy and conventional method is also a promising prospect. Hyperthermia using single domain magnetic nanoparticles (MNPs) is intensively studied; MNPs absorb energy from an external source that is usually an alternating magnetic field (AMF).

{\color{black}
The central quantity in MNP based hyperthermia is the absorbed power. It is related to the specific absorption rate, SAR, or specific loss power, SLP through a normalization with the MNP mass. SAR determines the efficiency of power absorption per unit sample mass, whose knowledge is important to assess the chances of hyperthermia as clearly the uptake of MNPs is limited in the organism. The classical definition of SAR is based on calorimetric measurements and it is defined as:\cite{{GARAIO2014432},{0957-4484-26-1-015704},{WANG2013}}
\begin{equation}
\text{SAR}=\frac{c_\text{s}m_\text{s}}{m_\text{NP}}\frac{\text{d}T}{\text{d}t}\Big{|}_{t=0},
\label{SARc}
\end{equation} 
where $c_\text{s}$ and $m_\text{s}$ are the specific heat and the mass of the sample, respectively, $m_\text{NP}$ is the mass of the MNPs and $T$ is its temperature. For \textit{in vitro} conditions, the measurement of SAR is crucial for the design of novel materials with a capability for hyperthermia treatment. 
For \textit{in vivo} studies, the most important quantity is the temperature of the tissue itself which depends on the absorbed power or the SAR therefore its knowledge is required for designing the thermal dosage.

The conventional methods to determine the absorbed power in ferrites use either the measurement of magnetization curves\cite{{GARAIO2014432},{GARAIO20142511},{0957-4484-26-1-015704},{CONNORD2014},{CARREY2011}} or a more direct calorimetric method\cite{{GARAIO2014432},{0957-4484-26-1-015704},{WANG2013},{ESPINOSA20161002}}. The earlier involves an electromagnetic modeling of the irradiation circuit and also the accurate knowledge of the magnetic properties of the ferrite material for the given irradiation frequency and magnitude of magnetic field. In addition, this method requires a highly homogenous AMF in a well defined geometry that calls for oversized irradiating coils and therefore a low efficiency of the input power. The calorimetric method requires to embed a non-metallic thermometer into the ferrite material itself or in the surrounding tissue. However, it is difficult to implement either of these methods in \textit{in vivo} conditions and the accuracy of the modeling is limited.} {\color{black}In addition, the conventional calorimetric method suffers from the so-called non-adiabatic condition\cite{PankhurstSAR} as it deduces the dissipated power while attaining a finite temperature difference between the sample and its environment. However, heat loss through heat conduction, convection, radiative loss, or evaporation \cite{PankhurstSAR} strongly limits the accuracy of this method.}

Herein, we present a method which circumvents all these limitations as it allows to determine the power absorbed during ferrite based hyperthermia via a non-calorimetric method. It is based on monitoring of the quality factor, $Q$ of the irradiation circuit. We show that a change in $Q$ upon placing a ferrite sample in the resonator with respect to a suitable reference provides an accurate means of determining the absorbed power. In addition, the method works well for small sized irradiation coils, i.e. the required input power can be significantly lower than in conventional studies. We validate the method by a comparison of the absorbed power with the theoretically expected values and also by comparing the result on the absorbed power with the more standard calorimetric approach. 
We also studied the method for a birdcage coil. The importance of this type of resonator is that it allows homogeneous RF irradiation for animal models and is readily available in magnetic resonance imaging (MRI) instruments.

\section{Results and discussion}

\subsection{Contactless assessment of dissipated power}
The method to directly obtain the absorbed power in a hyperthermia relevant ferrite samples is based on measuring the change in the resonator quality factor (or $Q$) after the ferrite sample is introduced with respect to a reference situation. In our proof of concept approach, we consider a resonator filled with water as reference, however a more realistic study should involve an phantom which is filled with an appropriate artificial tissue emulating (ATE) material \cite{Gabriel1,Gabriel2}. In general, the resonator $Q$ is the ratio of the energy stored in the resonator in the form of electromagnetic field and the power dissipated during a time period of the electromagnetic oscillation. As such, $Q$ is an accurate measure of the power dissipated due to the different mechanisms, such as e.g. ohmic losses in the resonator material, dielectric losses in capacitive elements, radiation from an inductive element, or dissipation in a sample which is placed inside the resonator. A common method to the determine the resonator $Q$ is to measure the power reflected from the resonator as a function of the irradiation frequency, which typically yields a Lorentzian profile, whose full width at half maximum gives $\Delta f$ and $Q=f_0/\Delta f$, where $f_0$ is the resonator frequency\cite{GYURE2015}. 

{\color{black}This gives:
\begin{gather}
Q=2\pi f_0\frac{\text{Energy stored in resonator}}{\text{Power loss}}=\frac{f_0}{\Delta f}.
\end{gather}

This relation is also valid for a resonator, whose original $Q_0$, is reduced in the presence of a sample, to $Q_{\text{sample}}$. Conservation of energy results in a well-known relation\cite{PooleBook,chen2004microwave}:
\begin{gather}
Q_{\text{sample}}^{-1}=Q_0^{-1}+L,
\end{gather}
where $L$ is a dimensionless quantity and is proportional to the power absorbed in the sample.
}

This type of measurement inevitably requires the use of an external circuit (also known as coupling of the resonator to the environment) whose presence affects the measured $Q$ (Ref. \onlinecite{GYURE2015}). In view of this, the literature distinguishes the quality factor of an ideal, uncoupled resonator, or $Q_0$ and that of a coupled (or loaded) resonator, or $Q_{\text{L}}$ with $Q_{\text{L}}<Q_0$. A special case is when the resonator is critically coupled, i.e. it does not reflect power back to the source when it is irradiated on its resonance frequency. Critical coupling is the only well defined situation when $Q_0$ can be measured with precision as for this case $Q_{\text{L}}=Q_0/2$. In practice, a reflected power of about 1\% (or in other words -20 dB) or smaller of the incoming power is usually considered as a critically coupled situation and we will also employ this threshold herein. 

The power dissipated in the ferrite, $P_{\text{ferrite}}$ can be obtained by measuring the quality factor in the presence of the ferrite, $Q_{\text{ferrite}}$, and with the appropriate reference material, $Q_{\text{ref}}$. We emphasize that for both measurements a near critical coupling is required. We derive in the Supplementary Material that:

\begin{gather}
P_{\text{ferrite}}=P_{\text{input}} \left(1-\frac{Q_{\text{ferrite}}}{Q_{\text{ref}}}\right),
\label{loss_calculation}
\end{gather}
where $P_{\text{input}}$ is the power transmitted to the resonator. Eq. \eqref{loss_calculation} returns $P_{\text{ferrite}}=P_{\text{input}}/2$ when $Q_{\text{ferrite}}=Q_{\text{ref}}/2$, $P_{\text{ferrite}}\approx 0$ when $Q_{\text{ferrite}}\approx{Q_{\text{ref}}}$, and $P_{\text{ferrite}}\approx P_{\text{input}}$ when $Q_{\text{ferrite}}\ll{Q_{\text{ref}}}$ as expected. In principle, the radiative resonator loss could be slightly modified upon inserting the sample into the resonator but we neglect this effect. We validate Eq. \eqref{loss_calculation} further below by calorimetric measurements.

In the following, we discuss the utility of our method i) to measure the absorbed power in the presence of ferrite in a sample and ii) for the measurement of SAR. In the absence of non-linear absorption effects, the power dissipated in the ferrite is proportional to the square of the RF magnetic field, which is proportional to the electromagnetic energy stored in the resonator. The latter quantity is proportional to the quality factor. We thus obtain for the resonator quality factor in the presence of the ferrite:

\begin{gather}
\frac{1}{Q_{\text{ferrite}}}=\frac{1}{Q_{\text{ref}}}+k\cdot Q_{\text{ferrite}},
\label{LoadedResonatorQ}
\end{gather}
where factor $k$ in Eq. \eqref{LoadedResonatorQ} depends on the ferrite volume and absorption properties, i.e. on the SAR. We note that a common mistake when calculating resonator $Q$ in the presence of a lossy sample is to neglect the effect of the sample absorption itself, thus we often find $\frac{1}{Q_{\text{ferrite}}}=\frac{1}{Q_{\text{ref}}}+k\cdot Q_{\text{ref}}$, which is only valid when the sample loss is small. 
\begin{figure}[htp]
\begin{center}
\includegraphics[scale=.45]{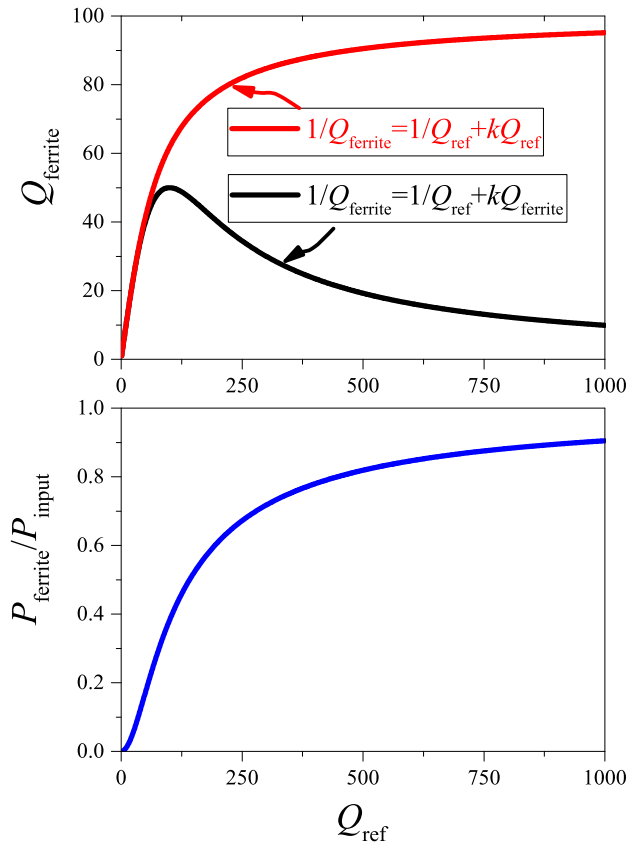}
\caption{Upper panel: The resulting resonator quality factor in the presence of the ferrite as a function of the reference quality factor, when calculated improperly (red curve) and self-consistently (black curve). Lower panel: the absorbed power as a function of the reference quality factor for a fixed value of $k=10^{-4}$.}
\label{Fig:Absorption_vs_Qreference}
\end{center}
\end{figure}

Fig. \ref{Fig:Absorption_vs_Qreference} shows the curve which is obtained using Eq. \eqref{LoadedResonatorQ} with $k=10^{-4}$ and also the absorbed power as a function of the resonator reference $Q$. Let $\sigma_a$ denote the standard deviation of $a=P_{\text{ferrite}}/P_{\text{input}}$. This quantity solely depends on the resonator $Q$ values with the ferrite and without, i.e. $a=a(Q_{\text{ref}},Q_{\text{ferrite}})$ according to Eq. \eqref{loss_calculation}. The error propagation theory yields:
\begin{gather}
\sigma_a^2=\left|\frac{\partial a}{\partial Q_{\text{ref}}}  \right|^2 \sigma^2\left( Q_{\text{ref}}\right)+\left|\frac{\partial a}{\partial Q_{\text{ferrite}}}  \right|^2 \sigma^2\left( Q_{\text{ferrite}}\right),
\end{gather}
since the two $Q$ measurements are uncorrelated. Given the simple form of $a=a(Q_{\text{ref}},Q_{\text{ferrite}})$, we obtain:
\begin{gather}
\frac{\sigma_a^2}{\left(1-a\right)^2}=\frac{\sigma^2\left( Q_{\text{ref}}\right)}{Q_{\text{ref}}^2}+\frac{\sigma^2\left( Q_{\text{ferrite}}\right)}{Q_{\text{ferrite}}^2}.
\label{AbsorbedPower_vs_Q_ErrorEstimate}
\end{gather}
This formula allows for a practical estimate of the absorbed power error in real life situations if the uncertainty of the $Q$ measurement is known. It is clear from Eq. \eqref{AbsorbedPower_vs_Q_ErrorEstimate}, that the error of the absorbed power, $\sigma_a$,  is smaller when $P_{\text{ferrite}}\rightarrow P_{\text{input}}$ (or $a\rightarrow 1$) for a given sample loss. Fig. \ref{Fig:Absorption_vs_Qreference}. demonstrates that in principle this could be achieved with the use of a high $Q$ reference resonator measurement. 
{\color{black}In practice, resonator $Q$ values are limited to a few hundred in the radio-frequency range (1-100 MHz) due to various effects including electric dissipation and radiative losses\cite{HARPEN1993}}.

\begin{figure}[htp]
\begin{center}
\includegraphics[scale=.45]{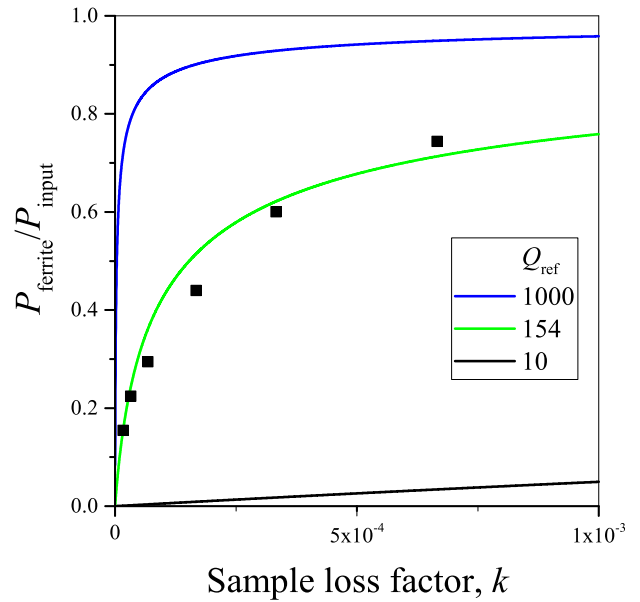}
\caption{The ratio of the power which is absorbed in the ferrite versus the exciting power as a function of the sample loss factor, $k$, which is proportional to the product of the SAR and sample mass for small $k$ values. Symbols show actual measurements for which the horizontal axis was scaled to match best the theoretical curve.}
\label{Fig:Absorption_vs_SAR_times_mass}
\end{center}
\end{figure}

The other area of interest for the proposed method is to determine the SAR for a standard hyperthermia candidate ferrite solution in laboratory conditions. Fig. \ref{Fig:Absorption_vs_SAR_times_mass} shows the variation of the absorbed power for three various values of the quality factor of the reference measurement as a function of varying sample mass (solid lines). The curves show a non-linear variation of $a=P_{\text{ferrite}}/P_{\text{input}}$ as a function of sample mass, which is the result of the non-linear Eq. \eqref{LoadedResonatorQ}. Solving Eq. \eqref{LoadedResonatorQ}. for the absorbed power ratio, $a$, we obtain:

\begin{gather}
a=1-\frac{-1+\sqrt{1+4 k Q_{\text{ref}}^2}}{2k Q_{\text{ref}}^2}.
\label{AbsorbedPowerRatio_solved}
\end{gather}
Remarkably, the curves given by Eq. \eqref{AbsorbedPowerRatio_solved} solely depend on the product $k Q_{\text{ref}}^2$, i.e. they are universal and fall on one another for different $Q_{\text{ref}}$ values and varying sample absorption.

Fig. \ref{Fig:Absorption_vs_SAR_times_mass}. also shows the absorbed power for a series of commercial Fe$_3$O$_4$ solutions with varying ferrite content (sample details are explained in the Methods section). The absorbed power was deduced from measuring the resonator quality factors for the samples according to Eq. \eqref{loss_calculation}. Clearly, the data follows well the expected non-linear curve, which attests the validity of the present approach. 

In practice, a series of measurements of $Q_{\text{ferrite}}$ with varying ferrite mass allows to determine the SAR parameter. The procedure is to first determine $Q_{\text{ref}}$, then measure $P_{\text{ferrite}}$ for different sample masses from $Q_{\text{ferrite}}$. This approach yields a constant SAR
when the sample little perturbs the resonator $Q$ as the curves  in Fig. \ref{Fig:Absorption_vs_SAR_times_mass}. start linearly with the sample mass since Eq. \eqref{AbsorbedPowerRatio_solved}, when expanded around $k=0$ yields:
\begin{gather}
P_{\text{ferrite}}=P_{\text{input}} \left[k Q_{\text{ref}}^2 +\mathcal{O} \left( k^2 Q_{\text{ref}}^4 \right)\right].
\label{AbsorbedPowerRatio_solved_approximated}
\end{gather}
Here, the SAR is identified as: $\text{SAR}=P_{\text{input}} k Q_{\text{ref}}^2/m_\text{NP}$, where $m_\text{NP}$ is in kilogram. It yields SAR in W/kg units. Here we emphasize that our method does not \emph{estimate} SAR but directly measures it as: $\text{SAR}=\frac{P_{\text{input}}}{m_\text{NP}}\left(1-\frac{Q_{\text{ferrite}}}{Q_{\text{ref}}} \right)$, which contains measured parameters only.

This approach to obtain a truly specific, i.e. a mass independent SAR, breaks down when the ferrite sample strongly affects the resonator $Q$. Then, one has to take into account the non-linearity of the absorbed power curves and the obtained SAR is reduced correspondingly. 

{\color{black}We note that the reference measurement (i.e. determination of $Q_{\text{ref}}$) plays a crucial role in order to obtain the extra power dissipated due the presence of the ferrite material. The best choice of reference measurement is a sample of the same material (preferably with the same heat capacity) except for the absence of the ferrite material itself.}

\subsection{Calorimetric validation of the method}

\begin{figure}[htp]
\begin{center}
\includegraphics[scale=.45]{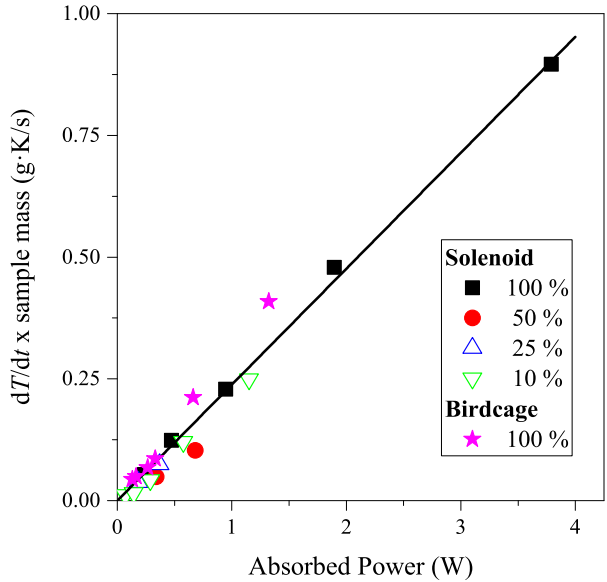}
\caption{The speed of warming per unit mass as determined from calorimetry for various sample concentrations as a function of the absorbed power as obtained from Eq. \eqref{loss_calculation} (100\% sample is the most concentrated as-received solution). The straight line is \emph{not a fit} to the data but has a slope of $\left(4.2\,\text{J/gK}\right)^{-1}$, i.e. it is a calculation using the specific heat of water. Note the excellent agreement between the data points and the calculation.}
\label{Fig:Absorption_vs_calorimetry}
\end{center}
\end{figure}

We further validate the present method by a comparison of the deduced absorbed power with direct calorimetric measurements. In Fig. \ref{Fig:Absorption_vs_calorimetry}. we show the speed of sample temperature warming multiplied by unit mass, $y=m_\text{s}\frac{\text{d}T}{\text{d}t}$ as a function of the absorbed power, $x=P_{\text{ferrite}}$. The quantity $y$ is obtained by monitoring the sample temperature whereas the absorbed power is obtained using Eq. \eqref{loss_calculation}. from a measurement of the resonator quality factors. In principle, the two quantities are related by:

\begin{gather}
m_\text{s}\frac{\text{d}T}{\text{d}t}=\frac{P_\text{ferrite}}{c_\text{s}}
\label{calorimetry_vs_Qmeas}
\end{gather} 
Fig. \ref{Fig:Absorption_vs_calorimetry}. shows a solid line: $y=x/c_\text{s}$ which is a calculated assuming $c_\text{s}=c_\text{water}\approx4.2\frac{\text{J}}{\text{gK}}$, i.e. that the sample consists of entirely water. It is worth noting that this approximation works well as the specific heat of the solution is much higher than that of the MNPs, which is the case for water. Although this is clearly an oversimplifying assumption, the experimental data points fall remarkably on this straight line.

{\color{black}Eq. \eqref{calorimetry_vs_Qmeas} highlights a major difference between the calorimetric measurement and the present method: the earlier requires to monitor the time dependence of sample heating and is therefore prone to non-adiabatic heating conditions, which are detailed in Ref. \onlinecite{PankhurstSAR}. Heat transfer effects from the sample to the environment through heat conduction, convection, or radiative losses greatly limit accuracy of calorimetry. However, the present method allows to monitor the power dissipated due to the presence of the ferrite \emph{in real-time}, i.e. even its variation with the sample temperature could be obtained.}

\subsection{Experimental validation on a birdcage resonator}

The so-called birdcage coil is a well-known type of RF resonators which are used extensively \cite{LEIFER199751,HARPEN1993,GIOVANNETTI2014,Lucano2016} in magnetic resonance imaging. It was first recommended by Hayes \textit{et al.} for such purposes in 1985 (Ref. \onlinecite{BirdcageDiscovery}). The primary reason for its utility in MRI is that a birdcage coil sustains a homogeneous RF magnetic field over a relatively large sample volume\cite{IBRAHIM2001}. The field direction is perpendicular to the coil axis when it is driven with a single RF signal however when driven with a quadrature RF signal, the bridcage coil produces a circularly rotating magnetic field, in contrast to e.g. a conventional solenoid or a surface coil, where the magnetic field is linearly polarized \cite{GIOVANNETTI2002}.  

\begin{figure}[htp]
\begin{center}
\includegraphics[scale=.45]{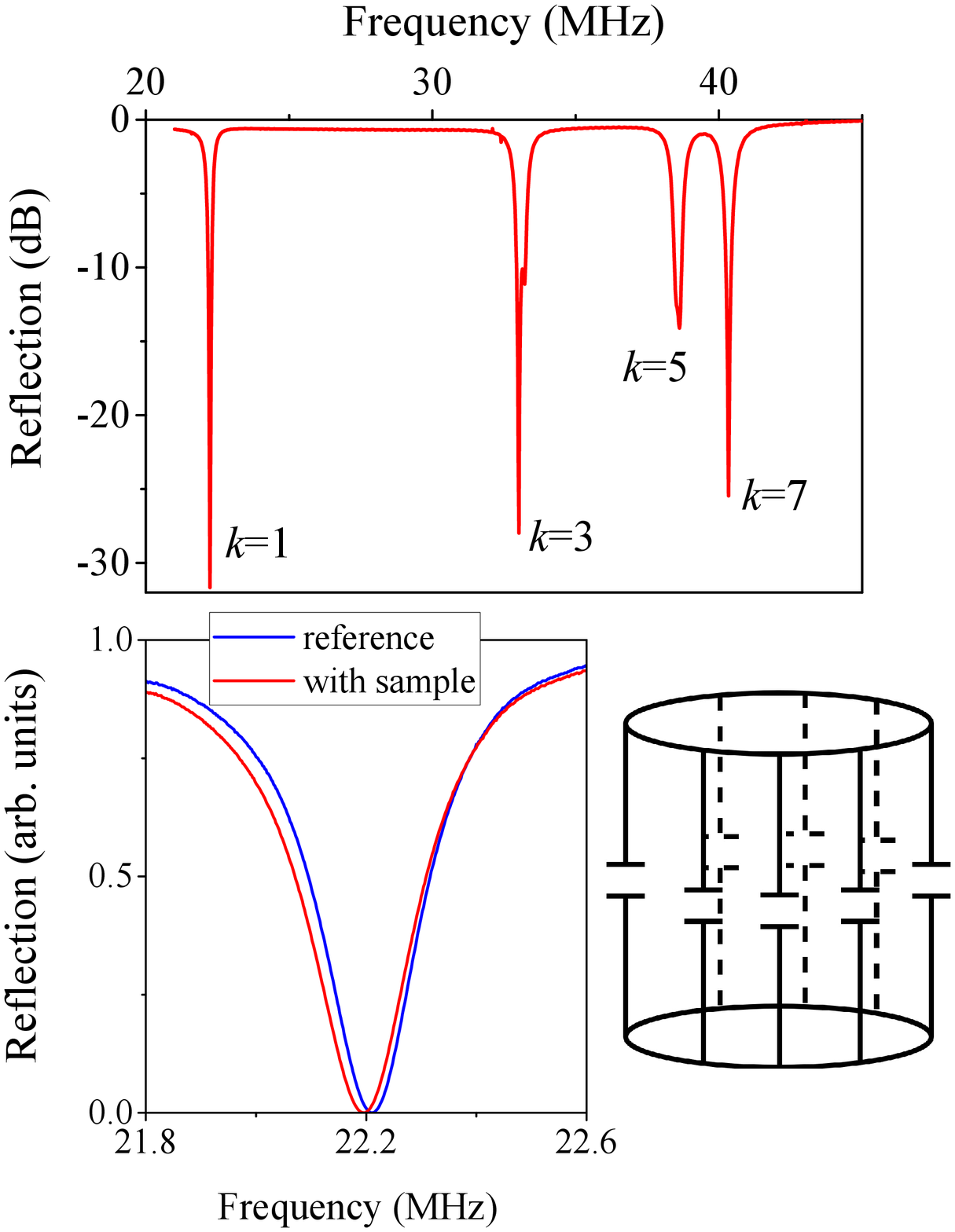}
\caption{Resonant modes of the 8-leg birdcage resonator as obtained from reflectometry (the magnitude of the $S$-parameter, $|S_{11}|$, is shown in dB units as a function of frequency). Note that the lowest, $k=1$ mode has the largest homogeneity, it is therefore used in MRI. A zoom-in is also shown for this mode with and without the sample (the reflection is shown on a linear scale). The overall schematics of the low-pass birdcage resonator is also provided.}
\label{Fig:Birdcage}
\end{center}
\end{figure}

The clear advantage of this resonator for thermotherapy purposes is that these are readily available as imaging coils and in fact it could be used for irradiation straight on. To test the applicability of our method for this type of resonator, we constructed a so-called low-pass birdcage resonator with 8 legs, which is shown in Fig. \ref{Fig:Birdcage}. The low-pass construction means that each leg is split into two by a capacitor of the same size (1 nF in our case). More details on the birdcage construction is given in the Supplementary Material. An 8-leg birdcage resonator is known to have 4 resonant modes (reflection curve is shown in Fig. \ref{Fig:Birdcage}.), of which the lowest frequency, $k=1$, mode sustains the most homogeneous RF magnetic field, therefore it can be used for MRI. The $Q$ factor of this mode is indeed sensitive for the presence of a ferrite sample as Fig. \ref{Fig:Birdcage}. demonstrates: the resonance curve is shifted and it is also slightly broadened. 

This type of coil is less sensitive to the same amount of ferrite sample than the solenoid due to the lower filling factor, $\eta$. For the presented study, the sample was as small as filling only 2\% of the birdcage volume, although the same volume would fill about 50\% of the solenoid volume. Correspondingly, the sample induced $Q$ factor change is much smaller. However, the most important property of our method is that it provides a sample or $\eta$ independent information on the absolute value of the absorbed power. This is clearly demonstrated in Fig. \ref{Fig:Absorption_vs_calorimetry}.: therein, we compare calorimetric results (or temperature change) with the absorbed power. The latter is calculated from the $Q$ factor change, using the method and the formula above. The result is robust and it shows that the dissipated power can be very accurately determined for the birdcage configuration when the filling factor is inevitably small.

\subsection{Limit of detection of the dissipated power}

We discussed above, that the statistical error of the dissipated power, $\sigma_a$ (where we defined above $a=P_{\text{ferrite}}/P_{\text{input}}$) depends on the error of $Q$ measurement. It was shown previously that the relative error of the $Q$ measurement, $\sigma(Q)/Q$, is independent of the $Q$ value and only depends on the applied method\cite{GYURE2015}. In our case, it is $\sigma(Q)/Q\approx 10^{-3}$. This, combined with our studies on the birdcage resonator, allows to give the limit of detection for the dissipated power for the given ferrite sample (Fe$_3$O$_4$) for the birdcage resonator. We employed a ferrite sample of 130 mg that gave $a=P_{\text{ferrite}}/P_{\text{input}}=0.030(1)$. 

According to Eq. \eqref{AbsorbedPower_vs_Q_ErrorEstimate}, we obtain $\sigma_a=(1-a)\times \sqrt{2}\cdot 10^{-3}$. This means that this sample amount gives a dissipated power which is 15 times larger than its error, i.e. $a=22\sigma_a$. As a result, for the present birdcage coil, a sample of about $6\,\text{mg}$ is the limit of detection for the absorbed power. {\color{black}This ferrite amount is enormous even for a practical \textit{in vitro} study. The amount of required ferrite could be reduced either with the use of a ferrite with an SAR larger than that of Fe$_3$O$_4$ or with the refinement of the sensitivity of the present method.} One has to consider that we employed a commercial Fe$_3$O$_4$ sample and magnetic nanoparticle materials with improved absorption, and as large as $\sim 30$ times larger SAR\cite{Bae2012}, are available. This could substantially reduce the limit of absorbed power detection of our method to sample amounts as low as 0.2 mg. \textit{In vivo} thermotherapy studies in mice employ a typical sample amount of 0.1-2 mg of magnetic nanoparticles \cite{ESPINOSA20161002,Heidari2016,Huang2013}. This means that our method provides a sensitivity of the absorbed power measurement which could be used for \textit{in vivo} thermotherapy studies. 

{\color{black}Our above estimate of the accuracy of the dissipated power measurement considers stochastic error sources only, and disregards systematic errors related to e.g. reproducibility when replacing the samples. However, the present method definitely overperforms conventional calorimetric studies when small absorbed power variations are to be monitored in-situ without replacing the sample. In such cases, calorimetry is known to have about $\pm5\%$ error in determining the dissipated power\cite{PankhurstSAR}.}

\section*{Conclusions}

We presented a non-calorimetric method which yields the absorbed power during nanomagnetic hyperthermia. This method, when combined with more conventional chemo- or radiotherapy, has growing significance in battling malignant tissues. It is based on the accurate measurement of the quality factor of resonators, that is used to sustain the irradiating radio frequency magnetic field. It represents an improvement over existing methods for similar purposes as these are either invasive or rely on an inaccurate modeling. {\color{black}The method allows to obtain the dissipated power in real-time and is not limited by the so-called non-adiabatic conditions of conventional calorimetric methods.} We validate our method by comparing the determined dissipated power, with data obtained with the more conventional calorimetric approach. The method allows for an alternative measurement of the specific absorption rate. We also show that besides conventional solenoids, the method performs well on birdcage coils; these are volume coils which sustain a very homogeneous RF field and are important in magnetic resonance imaging. We envisage that our method paves the way for the use of existing imaging hardware for RF based hyperthermia.

\section*{Methods} 

We used a commercial, water-based ferrite solution (Ferrotec EMG 705) which contains magnetite, Fe$_3$O$_4$. The producer supplied volume fraction is $3.6\,\%$ and its density is $1.19~\frac{\text{g}}{\text{cm}^3}$. We refer to the as-received most concentrated sample as 100\% sample. We prepared samples with a few different concentrations by diluting it with de-ionized water with a density of $1~\frac{\text{g}}{\text{cm}^3}$, the sample properties are provided in Table \ref{tab:mintak}. The samples were placed inside cylindric test-tubes with outer diameter of 5 mm and height of 30 mm.
\begin{figure}[htp]
\begin{center}
\includegraphics[scale=.6]{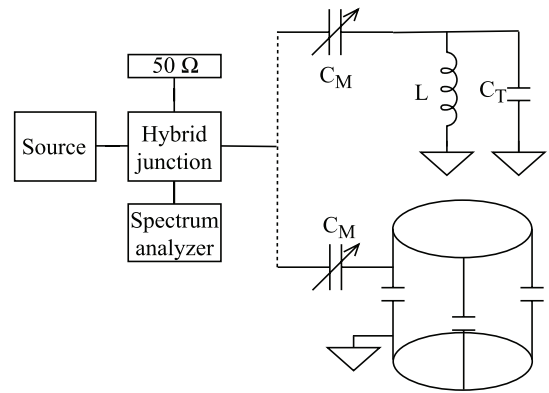}
\caption{The RF reflectometry setup. The hybrid junction divides the incoming RF power into 2 equal parts. One half is dissipated on the $50~\Omega$ resistor, the other half enters the resonant circuit. By trimming the tuning ($C_\text{T}$) and matching ($C_\text{M}$) capacitors, one can minimize the reflection for a certain frequency. The reflected power is measured using a broadband spectrum analyzer.}
\label{Fig:Setup}
\end{center}
\end{figure}

We studied two kinds of resonant circuits: a conventional tank circuit \cite{fukushima1981experimental} which consists of a solenoid for the test-tube studies and a so-called birdcage coil which is employed \cite{BirdcageDiscovery} in magnetic resonance imaging. High power/voltage trimmer capacitors (Voltronics Corp.) are used to match the circuits to 50 Ohm for efficient power transmission. The birdcage coil also contains fixed chip capacitors which set its resonant frequency. 
\begin{table}[htp]
		\begin{tabular}{|c|c|c|} \hline
			\textbf{dilution rate} & \textbf{magnetite mass (g)} & \textbf{sample mass (g)}\\ \hline
			1 (original) & 0.133 & 0.701\\\hline
			2 & 0.067 & 0.645\\\hline
			4 & 0.034 & 0.617\\\hline
			10 & 0.013 & 0.600\\\hline
		\end{tabular}
	\caption{The weight and dilutions of the investigated samples.}
	\label{tab:mintak}
\end{table}
 
Quality factors were determined using RF reflectometry with a 180 degree hybrid junction (ANZAC HH107) duplexer. Fig. \ref{Fig:Setup}. shows the RF reflection setup and we discuss further details of the RF reflectometry in the Supplementary Material. We match the circuits to minimal reflection using a scalar network analyzer (SignalHound, model SA124B for spectrum analyzer and TG124A tracking generator as source). To determine the resonator parameters accurately, we employ a swept signal source (Siglent SDG 1032) combined with a power detector (HP 8472B), which is connected to an oscilloscope (Tektronix TBS 1042). Lorentzian curves are fitted to determine 
the resonant frequency, $f_0$, and the linewidth, $\Delta f$ (which is the full width at half maximum, FWHM) of the resonance curve. The quality factor is obtained as $Q=f_0/\Delta f$. This also allows to average the reflected signal for a number of measurements, which enables to determine the mean and variation of the respective parameters.

The literature of resonators distinguishes \cite{chen2004microwave} the quality factor of the ideal, strongly undercoupled ($\beta \ll 1$) resonator, $Q_0$ and that of the coupled (or loaded) resonator, $Q_{\text{L}}$. When the resonator is critically coupled ($\beta=1$), i.e. there is no reflection from it ($S_{11}=0$), $Q_{\text{L}}=Q_0/2$. The losses in the resonator can be due to dielectric, ohmic, and radiative losses. The latter two dominates in our case as the capacitors have a high quality factor (about 1000). The loaded quality factor or $Q_{\text{L}}$ is typically between 20-100 for a RF circuit, as in our case. 

We use the terminology, $Q_{\text{ferrite}}$ and $Q_{\text{ref}}$, for the loaded quality factor of the resonator with and without the ferrite sample, respectively. This means that upon placing the sample under study into the resonator, the matching has to be readjusted to achieve zero reflection ($S_{11}=0$). The power which is incident on the resonator, $P_{\text{input}}$ is divided between the loss in the resonator and the loss in the sample as, $P_{\text{ferrite}}$, therefore we obtain for the latter quantity: 

\begin{gather}
P_{\text{ferrite}}=P_{\text{input}} \left(1-\frac{Q_{\text{ferrite}}}{Q_{\text{ref}}}\right).
\label{loss_calculation_methods}
\end{gather}
Eq. \eqref{loss_calculation} returns $P_{\text{ferrite}}=P_{\text{input}}/2$ when $Q_{\text{ferrite}}=Q_{\text{ref}}/2$, $P_{\text{ferrite}}\approx 0$ when $Q_{\text{ferrite}}\approx{Q_{\text{ref}}}$, and $P_{\text{ferrite}}\approx P_{\text{input}}$ when $Q_{\text{ferrite}}\ll{Q_{\text{ref}}}$ as expected. In principle, the radiative resonator loss could be slightly modified upon inserting the sample, but we neglect this effect.

The choice of a reference sample is crucial in our case. We found that the most appropriate reference measurement can be performed using a sample containing pure water only. The reason is that water itself can give rise to a slight change in the resonator properties.

We studied nanoparticle based hyperthermia around the 20-30 MHz frequency range. Most hyperthermia studies use a somewhat lower frequency. Our choice was motivated by the frequency range of MRI (which is usually 30-120 MHz) and the availability of well developed RF resonator techniques. In addition, we present a proof of concept i.e. the choice of frequency is somewhat arbitrary. We also note that an ongoing research focuses on the use of magnetic nanoparticles as an MRI contrast agent in addition to hyperthermia\cite{{ESTELRICH2014},{RAVICHANDRAN2017}}. 

We used a synthesized signal generator (Siglent SDG1025) followed by a power amplifier (Amplifier Research 306781) for irradiating the sample. The output power was calibrated with a power meter (Mini-Circuits PWR-SEN-6GHS). Calorimetry was performed using a optical thermometer (Luxtron FOT LAB Kit) with an accuracy of 0.01 K and temperature values were read out in every second. This allowed to determine $\frac{\text{d}T}{\text{d}t}$ in the beginning of a heating cycle well before temperature saturation sets in due to heat conduction toward the environment. This value is then directly proportional to the power absorbed in the solution containing the ferrites through: $P_{\text{absorbed}}=c_\text{s} m_\text{s} \frac{\text{d}T}{\text{d}t}$.

\section*{Acknowledgements}
Work supported by the Hungarian National Research, Development and Innovation Office (NKFIH) Grant Nrs. 2017-1.2.1-NKP-2017-
00001 and K11944.

\section*{Author Contributions}
GI performed the experiments and analyzed the data, Gy. T. provided the ferrite sample and initiated the nanoparticle based thermotherapy studies. BG and OS wrote the computer code which was used to obtain the data. BGM assisted the sample preparation and prepared some of the figures. FS suggested the presented method and outlined the overall composition of the paper. All authors contributed to writing of the manuscript.

\section*{Additional Information}
\textbf{Supplementary information} accompanies this paper at doi: {\color{red}to be filled by the publisher}.
\textbf{Competing interests:} The authors declare no competing interests.


\appendix
\newpage
\pagebreak
\clearpage

\section{Derivation of the expression for the power absorbed in the sample} 

We assume that the resonator which is loaded with the sample is critically coupled, i.e. maximum 1\% of the power can be reflected back to the source, or $S_{11}$ is smaller than -20 dB. This incoming power is thus nearly equal to the driving power of the source, i.e. $P_{\text{input}}\sim P_{\text{source}}$. In the absence of added ferrite, the resonator quality factor is $Q_{\text{ref}}$ and is filled with water in our simplified case, or with a phantom filled with an appropriate artificial tissue emulating material. In the presence of a ferrite material, the quality factor is lowered to $Q_{\text{ferrite}}$ due to the additional loss. The input power dissipation is divided between the resonator (this includes all kinds of losses also due to the solvent in which the ferrite is dissolved) plus the ferrite sample as:

\begin{gather}
P_{\text{input}}=P_{\text{resonator}}+P_{\text{ferrite}},
\end{gather}
where $P_{\text{resonator}}$ and $P_{\text{ferrite}}$ denote the power dissipated in the resonator and the additional ferrite, respectively. The ratio of the two terms can be obtained from the respective quality factors:

\begin{gather}
\frac{P_{\text{resonator}}}{P_{\text{ferrite}}}=\frac{Q_{\text{ref}}^{-1}}{Q_{\text{ferrite}}^{-1}-Q_{\text{ref}}^{-1}}.
\end{gather}
Solving these two equations yields:
\begin{gather}
P_{\text{ferrite}}=P_{\text{input}} \left(1-\frac{Q_{\text{ferrite}}}{Q_{\text{ref}}}\right),
\label{loss_calculation_SM}
\end{gather}

We also show in the main manuscript that the resonator quality factor in the presence of the ferrite, $Q_{\text{ferrite}}$ can be deduced as:
\begin{gather}
\frac{1}{Q_{\text{ferrite}}}=\frac{1}{Q_{\text{ref}}}+k\cdot Q_{\text{ferrite}},
\label{LoadedResonatorQ_SM}
\end{gather}
whose relevant solution is:
\begin{gather}
Q_{\text{ferrite}}=\frac{-1+\sqrt{1+4 k Q_{\text{ref}}^2}}{2k Q_{\text{ref}}}
\label{LoadedResonatorQ_solution_SM}
\end{gather}

This allows to obtain the absorbed power ratio, $a=P_{\text{ferrite}}/P_{\text{input}}=1-Q_{\text{ferrite}}/Q_{\text{ref}}$ as 
\begin{gather}
a=1-\frac{-1+\sqrt{1+4 k Q_{\text{ref}}^2}}{2k Q_{\text{ref}}^2}.
\end{gather}
This function is plotted in the main manuscript. It is clear that this function only depends on the product $k Q_{\text{ref}}^2$, therefore its shape is universal for the resonator loss problem.

\section{Details of the quality factor measurement}

\begin{figure}[htp]
\begin{center}
\includegraphics[scale=.23]{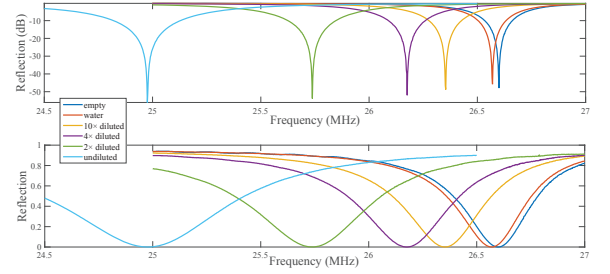}
\caption{The reflection curves for resonators which are empty or filled with samples. The upper panel shows the $S_{11}$ reflection parameter on log scale as obtained with the scalar network analyzer. The lower panel shows the reflected power shown on a linear scale. Critical coupling was achieved for all measurements.}
\label{Fig:Reflection}
\end{center}
\end{figure}

Fig. \ref{Fig:Reflection}. shows the results of reflectometry obtained using a scalar network analyzer and also the same type of data obtained with a power detector. The earlier allows for a more accurate determination of the quality of circuit matching, whereas the latter allows for a better determination of the resonator parameters, quality factor and resonant frequency. In the latter measurement, a computer control also allows for a repeated data acquisition which leads to a good estimate of the mean and variance of these parameters.

\section{Details of the calorimetric measurements} 

\begin{figure}[htp]
\begin{center}
\includegraphics[scale=.45]{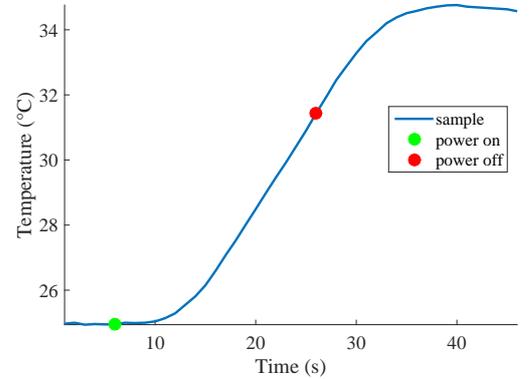}
\caption{Temperature increase of the sample (solid curve). Green and red circles indicate the start and the stop time of the RF irradiation.}
\label{Fig:Temp}
\end{center}
\end{figure}

Fig. \ref{Fig:Temp}. shows a typical time dependent heating curve using the RF irradiation. The apparently linear domain in $T(t)$ allows to determine the $\frac{\text{d}T}{\text{d}t}$ derivative using a linear fit onto the the steepest part of the curve. It is important to mention that there is a lag between the onset of temperature rise with respect to the power turn-on due to the thermal inertia of the sample. The warm up may also cause thermal circulations\cite{WANG2013} but this effect is neglected. The RF coil also warms up considerably as we do not employ a cooling. Its effect is, however, minimized by a good thermal isolation between the sample and the coil. In practice, an RF coil for hyperthermal treatment is made of copper tube with an appropriate cooling water flow\cite{GARAIO20142511}. When this effect become significant, it may also lead to an unwanted resonator detuning.

\section{Additional information on the birdcage resonator} 

\begin{figure}[htp]
\begin{center}
\includegraphics[scale=.6]{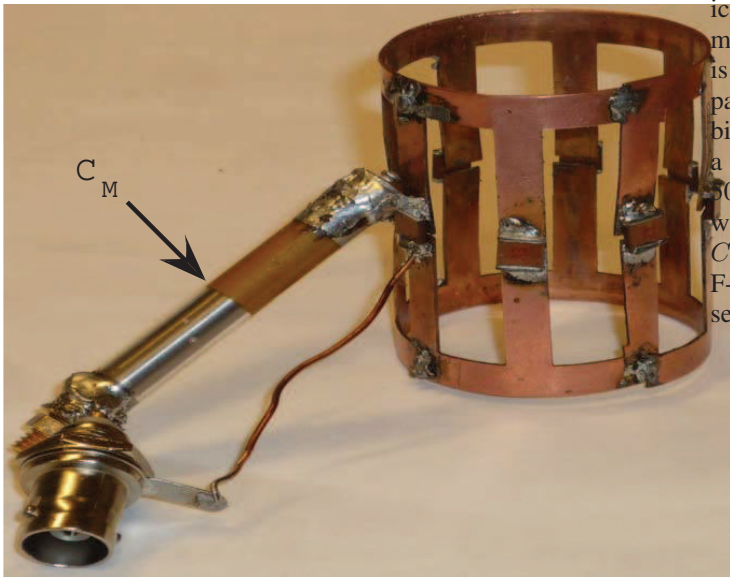}
\caption{Photograph of the 8-leg low-pass birdcage coil. A trimmer capacitor acts as circuit matching element.}
\label{Fig:Birdcage_photo}
\end{center}
\end{figure}

Fig. \ref{Fig:Birdcage_photo}. shows a photograph of the employed birdcage coil. Note the low-pass construction of the coil, i.e. that capacitors are in the middle of the legs. A trimmer capacitor (Voltronics Inc.) is soldered to the coil which serves for the circuit matching. It is important that the matching trimmer capacitor is soldered as close as possible to the two electrodes of a capacitor as otherwise no perfect matching can be achieved. The birdcage dimensions were designed to allow the irradiation of a laboratory mouse, it has a diameter of 30 mm and length of 50 mm. The coil body is made of copper stripes with 3 mm width and 1 mm thickness. The high-$Q$ RF capacitors have $C=1000\,\text{pF}$ (type CORNELL DUBILIER - MC22FA102J-F-CAP, 100V) with a variation of $\pm 5\,\%$ of which a few were selected with $\pm 1\,\%$ capacitance variation.

\end{document}